\documentclass[twocolumn]{aastex631}
\usepackage{amsmath}
\usepackage{graphicx}
\usepackage{placeins}
\usepackage{float}
\usepackage{txfonts}
\usepackage{hyperref}

\usepackage{lipsum}

\clearpage

\begin{document}

\title{Low-latency Forecasts of Kilonova Light Curves for Rubin and ZTF}

\author[0009-0008-8062-445X]{Natalya Pletskova}
\affiliation{Department of Physics, Drexel University, Philadelphia, PA 19104, USA}

\author{Niharika Sravan}
\affiliation{Department of Physics, Drexel University, Philadelphia, PA 19104, USA}

\author[0000-0002-9108-5059]{R. Weizmann Kiendrebeogo}
\affiliation{IRFU, CEA, Université Paris-Saclay, F-91191 Gif-sur-Yvette, France}
\affiliation{Laboratoire de Physique et de Chimie de l'Environnement, Université Joseph KI-ZERBO, Ouagadougou, Burkina Faso}
\affiliation{Artemis, Observatoire de la Côte d'Azur, Université Côte d'Azur, Boulevard de l'Observatoire, F-06304 Nice, France}
\affiliation{School of Physics and Astronomy, University of Minnesota, Minneapolis, Minnesota 55455, USA}

\author[0000-0002-8262-2924]{Michael W. Coughlin}
\affiliation{School of Physics and Astronomy, University of Minnesota, Minneapolis, Minnesota 55455, USA}

\author[0000-0002-2184-6430]{Derek Davis}
\affiliation{Senior Postdoctoral Scholar, California Institute of Technology, Pasadena, CA 91125, USA}

\author[0009-0008-9546-2035]{Andrew Toivonen}
\affiliation{School of Physics and Astronomy, University of Minnesota, Minneapolis, Minnesota 55455, USA}

\author[0009-0003-6181-4526]{Theophile Jegou du Laz}
\affiliation{Division of Physics, Mathematics and Astronomy, California Institute of Technology, Pasadena, CA 91125, USA}

\author[0000-0002-2184-6430]{Tomás Ahumada}
\affiliation{Division of Physics, Mathematics and Astronomy, California Institute of Technology, Pasadena, CA 91125, USA}

\author[0000-0002-4843-345X]{Tyler Barna}
\affiliation{School of Physics and Astronomy, University of Minnesota, Minneapolis, Minnesota 55455, USA}

\author[0000-0003-3367-3415]{George Helou}
\affiliation{Caltech/IPAC, 1200 E. California Blvd. Pasadena, CA 91125, USA}

\author[0000-0001-7062-9726]{Roger Smith}
\affiliation{Caltech Optical Observatories, California Institute of Technology, Pasadena, CA  91125}

\author[0000-0001-7648-4142]{Ben Rusholme}
\affiliation{IPAC, California Institute of Technology, 1200 E. California
             Blvd, Pasadena, CA 91125, USA}

\author[0000-0003-2451-5482]{Russ R. Laher}
\affiliation{IPAC, California Institute of Technology, 1200 E. California
             Blvd, Pasadena, CA 91125, USA}

\author[0000-0003-2242-0244]{Ashish A. Mahabal}
\affiliation{Division of Physics, Mathematics and Astronomy, California Institute of Technology, Pasadena, CA 91125, USA}
\affiliation{Center for Data Driven Discovery, California Institute of Technology, Pasadena, CA 91125, USA}

\begin{abstract}

Follow-up of gravitational-wave events by wide-field surveys is a crucial tool for the discovery of electromagnetic counterparts to gravitational wave sources, such as kilonovae. Machine learning tools can play an important role in aiding search efforts. We have developed a public tool to predict kilonova light curves using simulated low-latency alert data from the International Gravitational Wave Network during observing runs 4 (O4) and 5 (O5). It uses a bidirectional long-short-term memory (LSTM) model to forecast kilonova light curves from binary neutron star and neutron star-black hole mergers in the Zwicky Transient Facility (ZTF) and Rubin Observatory's Legacy Survey of Space and Time filters. The model achieves a test mean squared error (MSE) of 0.19 for ZTF filters and 0.22 for Rubin filters, calculated by averaging the squared error over all time steps, filters, and light curves in the test set. We verify the performance of the model against merger events followed-up by the ZTF partnership during O4a and O4b. We also analyze the effect of incorporating skymaps and constraints on physical features such as ejecta mass through a hybrid convolutional neural network and LSTM model. Using ejecta mass, the performance of the model improves to an MSE of 0.1. However, using full skymap information results in slightly lower model performance. Our models are publicly available and can help to add important information to help plan follow-up of candidate events discovered by current and next- generation public surveys.
\end{abstract}


\section{Introduction}\label{sec:intro}

Gravitational-wave (GW) interferometers provide direct observations of spacetime distortions produced by mergers of compact objects such as neutron stars and/or black holes, as well as other energetic cosmic phenomena. Mergers containing at least one neutron star are expected to produce transient electromagnetic (EM) counterparts. The emission can span the entire EM spectrum, from prompt gamma-ray bursts (GRBs) and their afterglows with signatures from gamma rays to the radio, to the more ubiquitous kilonovae (KNe) with signatures in the optical and near-infrared \citep{Metzger}. On August 17, 2017, the binary neutron star (BNS) merger GW170817 \citep{AbEA2017b} was detected across the full EM spectrum: as the short gamma-ray burst GRB 170817A \citep{AbEA2017c, Goldstein2017, Savchenko2017a} and as the kilonova AT 2017gfo \citep{ Abbott2017a, Arcavi2017, Chornock2017, Coulter2017, 2017Cowperthwaite, 2017Diaz, 2017Drout, 2017Fong, 2017Gall, 2017Hu, 2017Kasliwal, 2017Lipunov, 2017McCully, 2017Nicholl, 2017Pian, 2017Shappee, Smartt2017, 2017Soares-Santos, 2017Tanvir, Utsumi, Valenti, Villar, Pozanenko}. Since then, nine additional compact merger events have been identified \citep{ 2019Coughlin, Hosseinzadeh, Kuin2019GCN, 2019Song, 2020aAbbott, 2020aAntier, 2020Ab, 2020Gompertz, 2020S, 2021A, 2021Anand, 2022Chattopadhyay, 2022Wang}. However, further EM counterpart discoveries have remained elusive.

The search for EM counterparts to GW sources presents several challenges. These rare and faint events fade rapidly within a few days, making rapid identification crucial \citep{Metzger,Smartt2017, Kasliwal2020, Ahumada2021}. The complexity is increased by the large sky regions associated with GW alerts, which frequently span hundreds to thousands of square degrees \citep{Petrov2022}. In addition, the increased sensitivity of GW detectors and improved data reduction pipelines have resulted in an exponential increase in the discovery rate of compact merger events \citep{Kiendrebeogo23}. Despite the growing rate of GW detections, observational follow-up efforts remain constrained by limited telescope time, field of view, and weather or scheduling constraints.

The Zwicky Transient Facility (ZTF), a collaborative public–private survey of the Northern sky, uses \textit{g}, \textit{r}, and \textit{i} filters and a wide 47 $deg^2$ field-of-view imager on the Palomar 48-inch telescope. ZTF has demonstrated the value of wide-field optical surveys in discovering and monitoring transients \citep{Bulla2019, 2019Graham, 2019Masci, 2020Dekany}. Looking ahead, the advent of the Vera C. Rubin Observatory conducting the Legacy Survey of Space and Time (LSST) will significantly increase the transient discovery rate \citep{Ivezic2019}. While this offers new opportunities for kilonova detection, it also makes the search more challenging \citep{Andrade2025}. Light curves from the Wide Fast Deep (WFD) survey, consisting of $\sim $ 90\% of the survey time \citep{Ivezic2019}, are expected to be too sparse to reliably identify KNe. The most promising approach to improving the potential for the discovery of LSST KN is through a well-designed Target of Opportunity (ToO) program. Strategies distilled using community input prioritize ToO observations of well-localized, high-significance BNS and NSBH mergers, using selection criteria such as a FAR $<$ 1 per year, a 90\% credible region size ($\Omega$) $<$ 500 deg$^2$, and high probabilities of mass ejected during the merger (HasNS $\geq$ 0.5 and HasRemnant $\geq$ 0.5). Based on realistic trigger rates and observing constraints, the plan would result in approximately 60–70 ToO activations over LSST’s 10-year survey, all fitting within the 3\% observing time allocated to ToO efforts \citep{Andreoni2024}. However, this number is falling as we continue to see fewer BNS events than predicted, suggesting the rate might be lower. At the same time, few ground-based telescopes can follow-up the KNe discovered by Rubin. These include the Keck telescopes \citep{Keck}, Gemini Observatory \citep{Gemini1998}, the Very Large Telescope (VLT) \citep{ESO}, Subaru Telescope \citep{Subaru1986}, the Extremely Large Telescope (ELT) \citep{ELT2023}, the Thirty Meter Telescope (TMT) \citep{TMT2013}, and the Magellan Telescopes \citep{Magellan2003}. Spectrocopic follow-up will only be possible for the brightest sources given the rapid evolution timescales of KNe. There is a need for tools to facilitate the identification of KNe in LSST alert streams to maximize its discovery potential.

Machine learning (ML) tools can help provide effective solutions for processing and identifying astronomical transients, including KNe. An example is SuperNNova, a deep learning model that uses neural networks to classify and predict supernova light curves using both simulated and real observational data \citep[e.g.,][]{DASH, 2024Berbel}. Recent studies have extended ML techniques to KN discovery and classification, including a temporal convolutional neural network (CNN) trained on early-time photometry and contextual information to identify KNe within GW localization regions \citep{Chatterjee2022}, an ML classifier designed to detect fast transients such as KNe in the ZTF public alert stream \citep{Biswas2024}, and a likelihood-free inference framework that uses a neural network–based embedding to estimate KN parameters using light curves \citep{Desai2024}. In addition, recent work has improved KN light curve modeling by incorporating uncertainties in the neutron star equation of state (EoS). By marginalizing over a range of candidate EoS models, simulations can produce robust predictions of ejecta masses and multiband light curves \citep{2024Toivonen}.

The International Gravitational Wave Network (IGWN) broadcasts real-time alerts of its discoveries to help astronomers find electromagnetic counterparts. The data in these alert packets include machine learning probabilities that provide additional insight into the properties of the merger \citep[e.g.,][]{Chatterjee2020}. Groups often use selection criteria using alert data to manage the alert volume and determine which events are promising and warrant further investigation. These criteria include thresholds on the probability that a signal is astrophysical ($P_{\text{Astro}}$) and the False Alarm Rate (FAR). However, these cuts do not take into account whether a kilonova would be observable under the survey cadence. For example, a low FAR event could produce a bright KN, which may be culled by selection cuts (see, e.g. Figure 12). A more effective approach would use the alert data to estimate the potential brightness of a kilonova, its light evolution over time, and the scheduled survey observations \citep{O4a, 2025Fulton}.

In this paper, we develop a tool to aid in the kilonova discovery and follow-up planning process. It uses low-latency alert data from the IGWN to forecast the expected KN characteristics in ZTF and Rubin filters. This tool could help offer insight on promising candidates and plan follow-up based on forecasted observability. This paper is organized as follows. Section 2 provides an overview of the training dataset for BNS and NSBH merger events. Section 3 describes the bidirectional long short-term memory (LSTM) model used to forecast kilonova light curves. Section 4.1 summarizes the model performance for ZTF filters, while Section 4.2 presents the model performance for Rubin filters. Section 4.3 explores the role of skymaps and physical features such as ejecta mass in the prediction of kilonova light curves using a hybrid convolutional neural network (CNN)-LSTM model. Section 4.4 evaluates how well the model predictions align with candidates followed up during IGWN observing runs O4a and b. We conclude in Section 5. Our models are publicly available on GitHub at \url{https://github.com/np699/Forecast-KN-LC.git}.

\section{Training Data}

Between 2015 and 2020, during the first three observing runs, the IGWN detected 2 BNS mergers and 4 NSBH mergers \citep{2023Abbott}. Given the limited number of observed events, simulated data is crucial for the development of ML models to support KN discovery and inference. Additionally, the ongoing fourth observing run (O4) has produced a growing number of public low-latency alerts, which provide new opportunities to test and refine these models in near real-time.

We use simulations of the observation scenarios described in the IGWN User Guide\footnote{\url{https://emfollow.docs.ligo.org/userguide/capabilities.html}}, which are publicly available to the researchers community \citep{Kiendrebeogo23}. The data set includes 17,009 simulated BNS events and 3,148 simulated NSBH events\footnote{\url{https://zenodo.org/records/12696721}}, all of which passed the signal-to-noise ratio (SNR) threshold of 8 for IGWN detection during observing runs 4 and 5. These events are selected from a larger set of 20 million compact binary coalescences (CBCs) used as injections. The mass threshold distinguishing neutron stars (NSs) from black holes (BHs) is set at 3 solar masses. However, for our work, we use a subset of the data consisting of 7,985 simulated BNS events and 92 simulated NSBH events from the O4 run to train models on ZTF filters, and 4,223 simulated BNS events and 82 simulated NSBH events from the O5 run to train on Rubin filters. These numbers are smaller than those of other simulations because we used a custom setup when generating the data. Specifically, we set $\alpha = 0$, $\text{ratio}_\zeta = 0.3$, and $\text{ratio}_\epsilon = 0$, which means that we turn off some sources of ejecta (fallback and wind) and only allow 30\% of the disk mass to be ejected. As a result, fewer systems produce enough ejecta to generate light curves. We adopt this set-up to enable a direct comparison with the ejecta mass predictions from \citet{2024Toivonen} work and to ensure consistency in the training set for our ML models.

To generate sky maps for simulated BNS and NSBH mergers, we use Bayestar \citep{Singer2016}, extracting key parameters such as the sky localization area enclosing 90\% of the probability mass (area(90)). The source distance is taken directly from the injection metadata. Since simulations only provide SNR, whereas public alerts from IGWN include FAR and $P_{\text{Astro}}$ ($P_{\text{BNS}}$+$P_\text{NSBH}$+$P_\text{BBH}$), we establish a mapping between SNR and FAR by fitting a large number of injections of BNS mergers from the Gravitational-Wave Transient Catalog 3 (GWTC-3) released by the LIGO, Virgo, and KAGRA Collaborations\footnote{\url{https://zenodo.org/records/7890437}}. This relationship is determined using the median of individual injections and is further refined with a rolling median; see figure 1, yielding:

\begin{equation}
\log_{10}(\text{IFAR}) = 2.357\times \text{SNR} - 20.198
\end{equation}
where IFAR is the inverse FAR.

\begin{figure}
    \centering
    \includegraphics[width=0.45\textwidth]{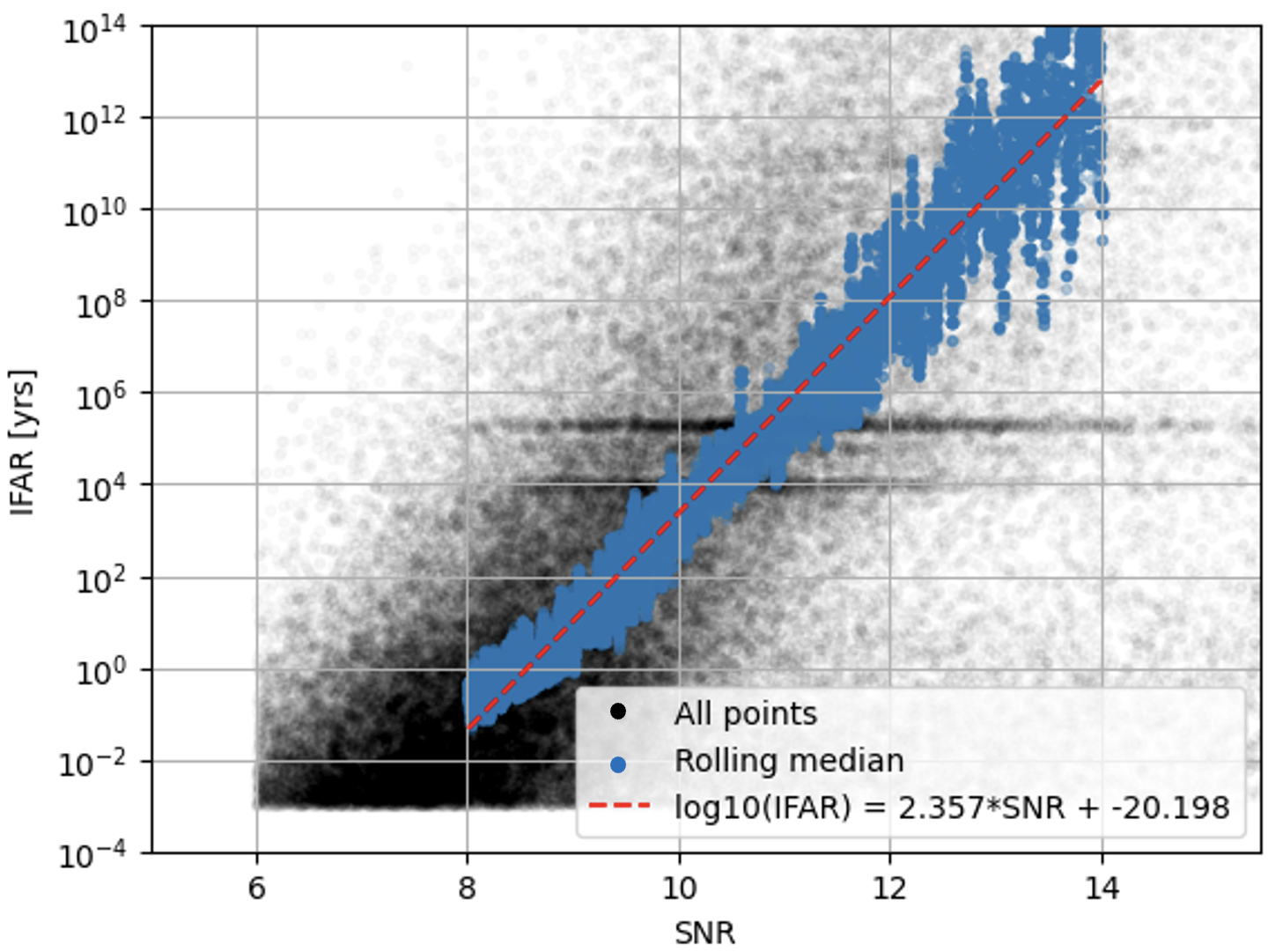}
    \caption{Mapping the signal-to-noise ratio to inverse false alarm rate using a large population of binary neutron stars injections.}
    \label{fig:lumfunc}
\end{figure}

$P_{\text{Astro}}$ is defined as:

\begin{equation}
P_{\text{Astro}} = 1 - P_{\text{Terrestial}},
\end{equation}
where $P_{\text{Terrestial}}$ represents the probability that a detected signal comes from terrestrial noise rather than an astrophysical source. To calculate $P_{\text{Terrestial}}$, first we compute the astrophysical event rate at the observed SNR for a given event as:

\begin{equation}
R_{\text{astro}} = R_\text{CBC} \times \left(\frac{SNR_\text{thresh}}{SNR}\right) ^3
\end{equation}
Using Eq. 1, this is:

\begin{equation}
R_{\text{astro}} = R_\text{CBC} \times \left(\frac{\frac{1}{2.357}(\log_{10}(\text{$IFAR_{\rm threshold}$}) + 20.198)}{\frac{1}{2.357} \left(\log_{10}(\text{IFAR}) + 20.198 \right)}\right) ^3
\end{equation}

Finally, 

\begin{equation}
P_{\text{Terrestial}}  = \frac {R_{\text{noise}}} {R_{\text{astro}} +R_{\text{noise}}},
\end{equation}
where $R_\text{noise}$ is $\frac{4}{IFAR}$. Here we are assuming 4 independent discovery pipelines running \citep{2023Abbott}. We also assume the IFAR threshold is 1 yr and $R_\text{CBC}$ is 35 per year per Gpc$^3$, consistent with estimates from the O3 observing run \citep{2023Abbott}. The O4 observing run probably yields a higher rate due to advancements in sensitivity and detection pipelines, with final numbers still under evaluation.

We also compute probabilities related to electromagnetic-bright (EM-bright)\footnote{\url{https://git.ligo.org/emfollow/em-properties/em-bright}} classifications: HasNS, HasRemnant, and HasMassGap (\citealt{Chatterjee2020}), which represent the probability of the detected strain signal to have arisen from a merger having at least one NS, a non-zero ejecta mass, and at least one constituent NS in the mass range of 3-5 solar masses, respectively. Finally, we use the Nuclear Multi-Messenger Astronomy (NMMA)\footnote{\url{https://nuclear-multimessenger-astronomy.github.io/nmma/fitting.html}} framework \citep{Pang2023}, which uses POSSIS models (\citealt{Bulla2019,Diet2020}), to generate light curves for every simulated CBC. This model estimates the kilonova emission based on the properties of the merging binary system. It uses fitting formulae, developed from detailed simulations, to translate inputs such as the masses of the binary components, how easily they are deformed by gravity (tidal deformability), and their spins into estimates of the material ejected during the merger \citep{Dietrich2020, 2018Foucart}. This includes both fast-moving dynamical ejecta and slower-moving wind ejecta from the leftover disk. These ejecta properties are then used to simulate how the kilonova would appear from different viewing angles using POSSIS models. This model accounts for a number of important factors that affect KN emission, such as the opening angle, the inclination angle, the dynamical ejecta, and the wind ejecta. The resulting dataset is divided into 70\% for training, 15\% for validation, and 15\% for testing.

\section{Model}

\begin{figure*}[t]
    \centering
     \includegraphics[width=0.98\textwidth]{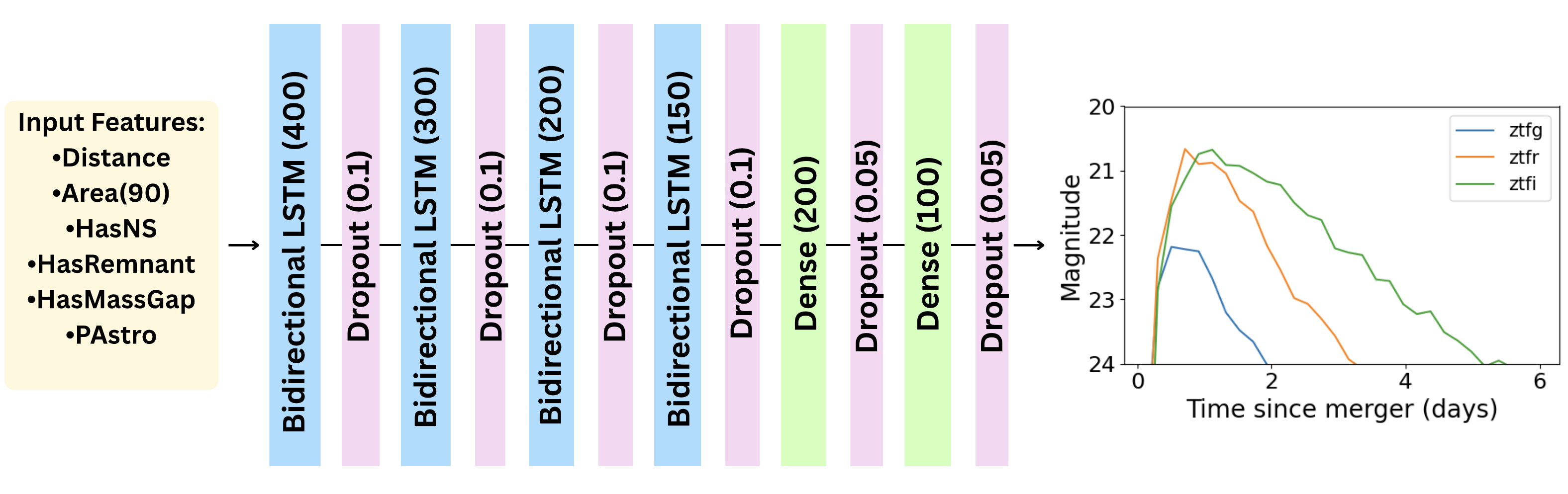}
    \caption{Architecture of the bidirectional LSTM model used for light curve predictions.}
    \label{fig:corner}
\end{figure*}

\begin{table}
\centering
\renewcommand{\arraystretch}{1} 
\begin{tabular}{|c|c|}
\hline
\multicolumn{2}{|c|}{\textbf{Input Features}} \\ 
\hline
\multicolumn{2}{|c|}{Distance, area(90), HasNS, HasRemnant, HasMassGap, $P_{\text{Astro}}$} \\ 
\hline
\multicolumn{2}{|c|}{\textbf{Bidirectional LSTM Layer 1}} \\ 
\hline
Units & 400 \\ 
\hline
Activation Function & ReLU \\ 
\hline
Return Sequences & True \\ 
\hline
Regularization (L1, L2) &  L1 = $10^{-5}$, L2 = $10^{-4}$ \\ 
\hline
Kernel Constraint & max\_norm(3.0) \\ 
\hline

\multicolumn{2}{|c|}{\textbf{Bidirectional LSTM Layer 2}} \\ 
\hline
Units & 300 \\ 
\hline
Activation Function & ReLU \\ 
\hline
Return Sequences & False \\ 
\hline
Regularization (L1, L2) &  L1 = $10^{-5}$, L2 = $10^{-4}$ \\ 
\hline
Kernel Constraint & max\_norm(3.0) \\ 

\hline
\multicolumn{2}{|c|}{\textbf{Bidirectional LSTM Layer 3}} \\ 
\hline
Units & 200 \\ 
\hline
Activation Function & ReLU \\ 
\hline
Return Sequences & False \\ 
\hline
Regularization (L1, L2) & L1 = $10^{-5}$, L2 = $10^{-4}$ \\ 
\hline
Kernel Constraint & max\_norm(3.0) \\ 
\hline

\multicolumn{2}{|c|}{\textbf{Bidirectional LSTM Layer 4}} \\ 
\hline
Units & 150 \\ 
\hline
Activation Function & ReLU \\ 
\hline
Return Sequences & False \\ 
\hline
Regularization (L1, L2) & L1 = $10^{-5}$, L2 = $10^{-4}$ \\ 
\hline
Kernel Constraint & max\_norm(3.0) \\ 
\hline

\multicolumn{2}{|c|}{\textbf{Dropout}} \\ 
\hline
Dropout & 0.1 after each LSTM layer \\ 
\hline

\multicolumn{2}{|c|}{\textbf{Fully Connected Dense Layer 1}} \\ 
\hline
Units & 200 \\ 
\hline
Activation Function & ReLU \\ 
\hline
Regularization (L1, L2) & L1 = 0.001, L2 = 0.001 \\ 
\hline
Dropout & 0.05 \\ 
\hline

\multicolumn{2}{|c|}{\textbf{Fully Connected Dense Layer 2}} \\ 
\hline
Units & 100 \\ 
\hline
Activation Function & ReLU \\ 
\hline
Regularization (L1, L2) & L1 = 0.001, L2 = 0.001 \\ 
\hline
Dropout & 0.05 \\ 
\hline

\multicolumn{2}{|c|}{\textbf{Output Layer}} \\ 
\hline
Units & 90 (30 steps within 6 days and 3 ZTF filters)\\ 
\hline
Activation Function & Linear \\ 
\hline

\end{tabular}
\caption{Architecture of our bidirectional LSTM model.}
\label{tab:hyperparameters}
\end{table}

We train an ML model to forecast kilonova light curves using low-latency alert data from IGWN. The model employs a bidirectional LSTM network as our fiducial model, which is well suited for working with sequential data, such as time series.

The model architecture is shown in Figure 2. Table 1 specifies the model configuration in more detail. To capture complex dependencies, the model includes four bidirectional LSTM layers with 400, 300, 200, and 150 neurons, respectively. We have adopted bidirectional LSTMs because of their ability to process sequential input in both forward and backward directions, allowing the model to identify patterns and relationships between both future and previous observations \citep{Schuster1997}. Dropout \citep{srivastava2014dropout} is used for regularization, with a rate of 0.1 after each bidirectional LSTM layer. Following the LSTM layers, the model incorporates two fully connected dense layers with 200 and 150 units before the final output layer. A final dense output layer with 90 units is used to map the processed features to the prediction of the KN light curve across three ZTF filters.

We use a grid search to tune hyperparameters such as batch size, regularization strength, learning rate, dropout rates, and the number of units in each bidirectional LSTM layer using a held out validation set. The best fit hyperparameters are listed in Table 2. We use MC dropout (with value 0.1) to estimate mean and uncertainity of forecasted light curves. In addition, we use RobustScaler to maintain uniformity in scale and measurement units throughout the training data.

\begin{table}[ht]
\centering
\begin{tabular}{|c|c|}
\hline
\textbf{Hyperparameter} & \textbf{Value} \\ \hline
Optimizer & Adam \\ \hline
Loss Function & Mean Square Error \\ \hline
Learning Rate (LR) & 0.0003 \\ \hline
Patience & 5 \\ \hline
Reduction Factor & 0.5 \\ \hline
Minimum LR & $10^{-6}$ \\ \hline
Batch Size & 64 \\ \hline
Epochs & 300 \\ \hline
\end{tabular}
\caption{Best fit hyperparameters of our bidirectional LSTM model.}
\end{table}

\section{Results}

\subsection{ZTF}

Our LSTM model is evaluated using the MSE loss function, which is ideal for regression tasks as it reduces the average squared differences between the predicted and actual light curves. In addition, we use the $R^2$ evaluation function to have additional verification of our performance. The value of $R^2$ ranges from 0 to 1, where $R^2$ = 1 indicates that the model explains all the changes in the dependent variable, $R^2$ = 0 suggests that the model explains none of the variance and negative values $R^2$ can occur if the model performs worse than the mean.

We obtain a mean squared error (MSE) of 0.19 and an $R^2$ score of 0.82 on the O4 test data across the ZTF \textit{g}-, \textit{r}-, and \textit{i}-band filters using our fiducial model for ZTF. We find that more than half of the test dataset (1,715 light curves) has an MSE below 0.2, indicating reliable predictions. Among them, 884 light curves achieve an MSE below 0.1, demonstrating high accuracy. However, some test cases exceed an MSE of 0.2, suggesting room for improvement. A typical light curve forecast is presented in Figure 3. The ground truth light curves are close to the 3-$\sigma$ uncertainty range across the three ZTF filters ( \textit{g}, \textit{r}, and \textit{i}). The best predicted light curve achieves an extremely low MSE of 0.0027, as shown in the left panel of Figure 4. In this case, the model predictions closely match the actual values. The right panel of Figure 4 shows the worst predicted light curve, with an MSE of 1.76. In this case, the model still captures the general shape of the light curve but makes larger mistakes in the predicted brightness. The true light curve is very faint and has an unusual shape, which probably made it harder for the model to predict accurately, especially since similar examples were rare in the training data (see also the discussion below and Figure 6). Interestingly, the model tends to underpredict the brightness in such cases instead of overpredict. While this means more uncertainty, it may actually be helpful for follow-up observations, since underestimating the brightness encourages observers to look deeper and avoid missing faint events.

Figure 5 shows the MSE for each ZTF filter as a function of time. Compared to the \textit{g} and \textit{i} filters, the \textit{r} filter performs slightly better during the earlier phases, having a lower MSE. After 2 days, all filters maintain a low MSE, suggesting a strong prediction accuracy. However, the \textit{i} filter performs best as shown by its reduced MSE compared to the \textit{g} and \textit{r} filters. 

\begin{figure*}[t]
    \centering
     \includegraphics[width=1\textwidth]{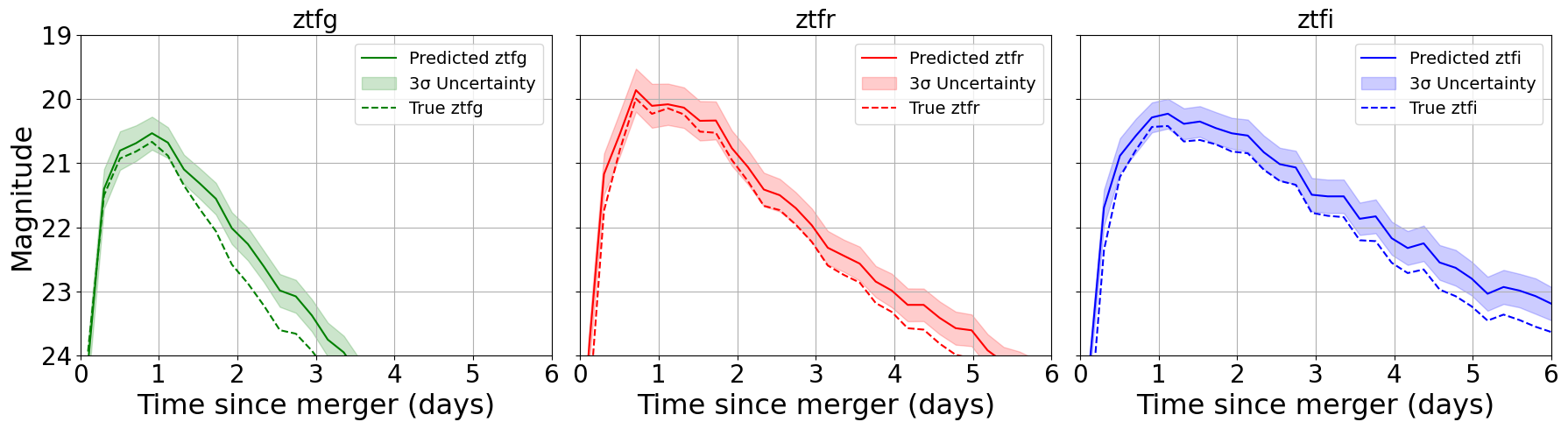}
    \caption{Predicted (solid lines) and ground truth (dashed lines) light curves in ZTF filters using our trained model for a random KN in our test set. Shaded regions show 3-$\sigma$ uncertainties.}
    \label{fig:corner}
\end{figure*}

\begin{figure*}
    \centering
    \includegraphics[width=0.45\textwidth]{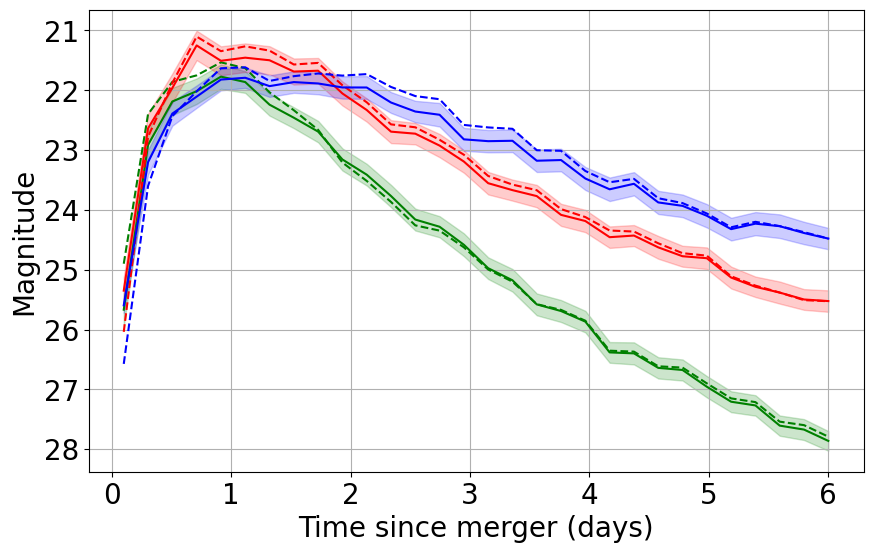}
    \includegraphics[width=0.45\textwidth]{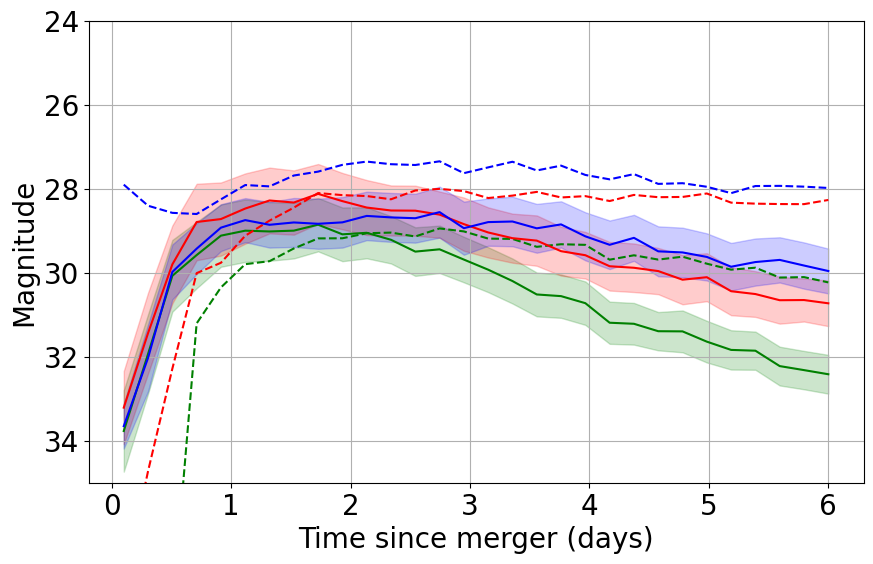}
    \caption{{\it Left:} The best-predicted light curve with an MSE of 0.0027. The ground truth (dashed lines) and predicted light curves (solid lines) match closely. {\it Right:} The worst predicted light curve, with an MSE of 1.76.}
    \label{fig:lumfunc}
\end{figure*}

\begin{figure}
    \centering
    \includegraphics[width=0.45\textwidth]{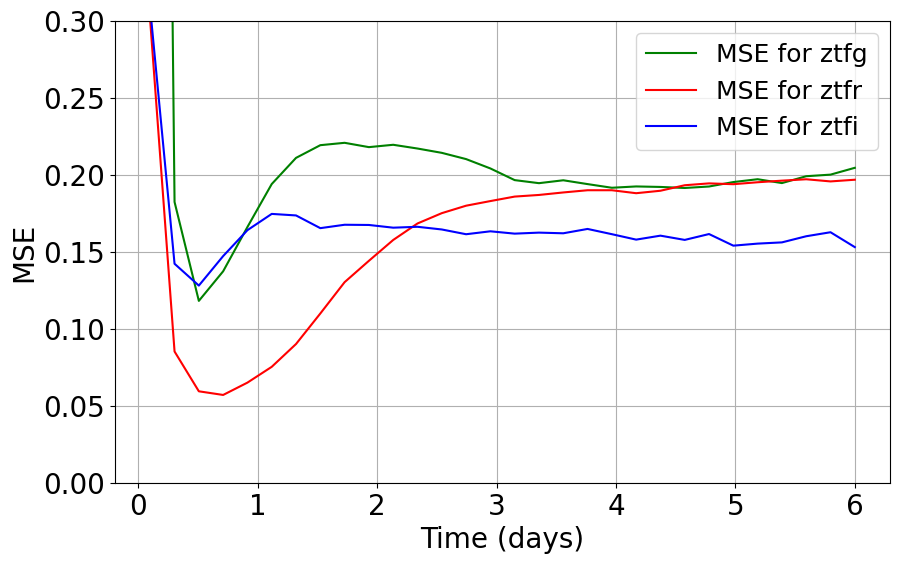}
    \caption{MSE as a function of KN phase for ZTF filters. Small MSE values indicate better model performance. Early time forecasts ($\sim$5 hours since merger) are unreliable. Forecasts in the r-filter are the best at early phases ($\lesssim$ 2 days).}
    \label{fig:lumfunc}
\end{figure}

\begin{figure*}[t]
    \centering
     \includegraphics[width=0.98\textwidth]{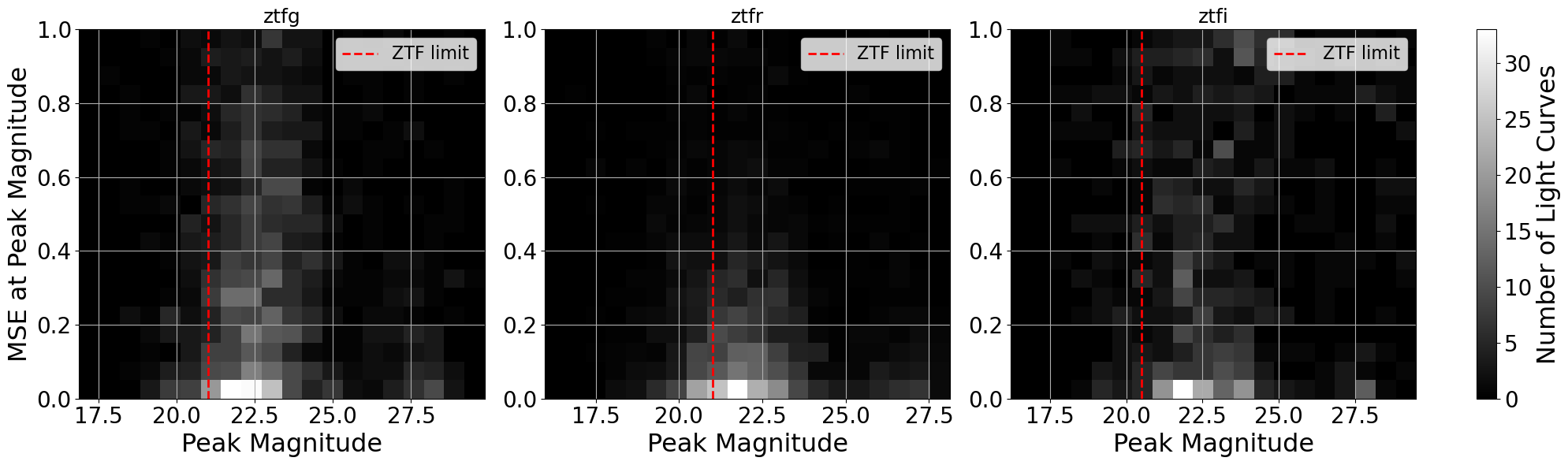}
    \caption{2D histograms showing the peak magnitudes of KN light curves in the training dataset versus the MSE of the forecasted light curves in ZTF \textit{g} (left), \textit{r} (middle), and \textit{i} (right) filters. The colorbar shows the number of training examples in each bin; larger counts are indicated by lighter colors. The model best performs for KNe with peak magnitudes (21–24) where the training dataset has highest number of examples. The red dashed lines mark the ZTF magnitude limits.}
    \label{fig:corner}
\end{figure*}

Figure 6 illustrates how model performance varies with peak magnitude and number of training examples in the ZTF  \textit{g}, \textit{r}, and \textit{i} bands. Across all filters, most of the light curves cluster around peak magnitudes of 21–24 with low MSE values, indicating strong predictive performance in this range. The \textit{r}-band exhibits the tightest clustering near zero MSE, suggesting that the model performs best in this filter. In contrast, the \textit{g}- and \textit{i}-bands show broader MSE distributions, particularly in the \textit{i}-band, where a higher fraction of events have elevated errors across the full range of magnitudes. The model performs worse for very bright events (peak magnitude $<$ 20), probably due to their under-representation in the training set. Similarly, performance declines for faint transients (peak magnitude $>$ 25), reflecting the limited detectability of distant or low-luminosity mergers in the injected population. This is because the observing scenarios used to generate the training dataset relied on injected populations that followed a uniform distribution in comoving volume. Since kilonova brightness is strongly dependent on distance, nearby (bright) events are rare in such a distribution. At the faint end, only compact binary CBCs with the highest SNR are detectable, typically those with larger total masses, which correlate with higher ejecta masses, and thus brighter KNe. The red dashed lines mark the ZTF magnitude limits \citep{ZTF}; notably, there are few training examples beyond these limits, which contributes to reduced model performance in those regions.

\subsection{Rubin model}

\begin{figure*}[t]
    \centering
     \includegraphics[width=1\textwidth]{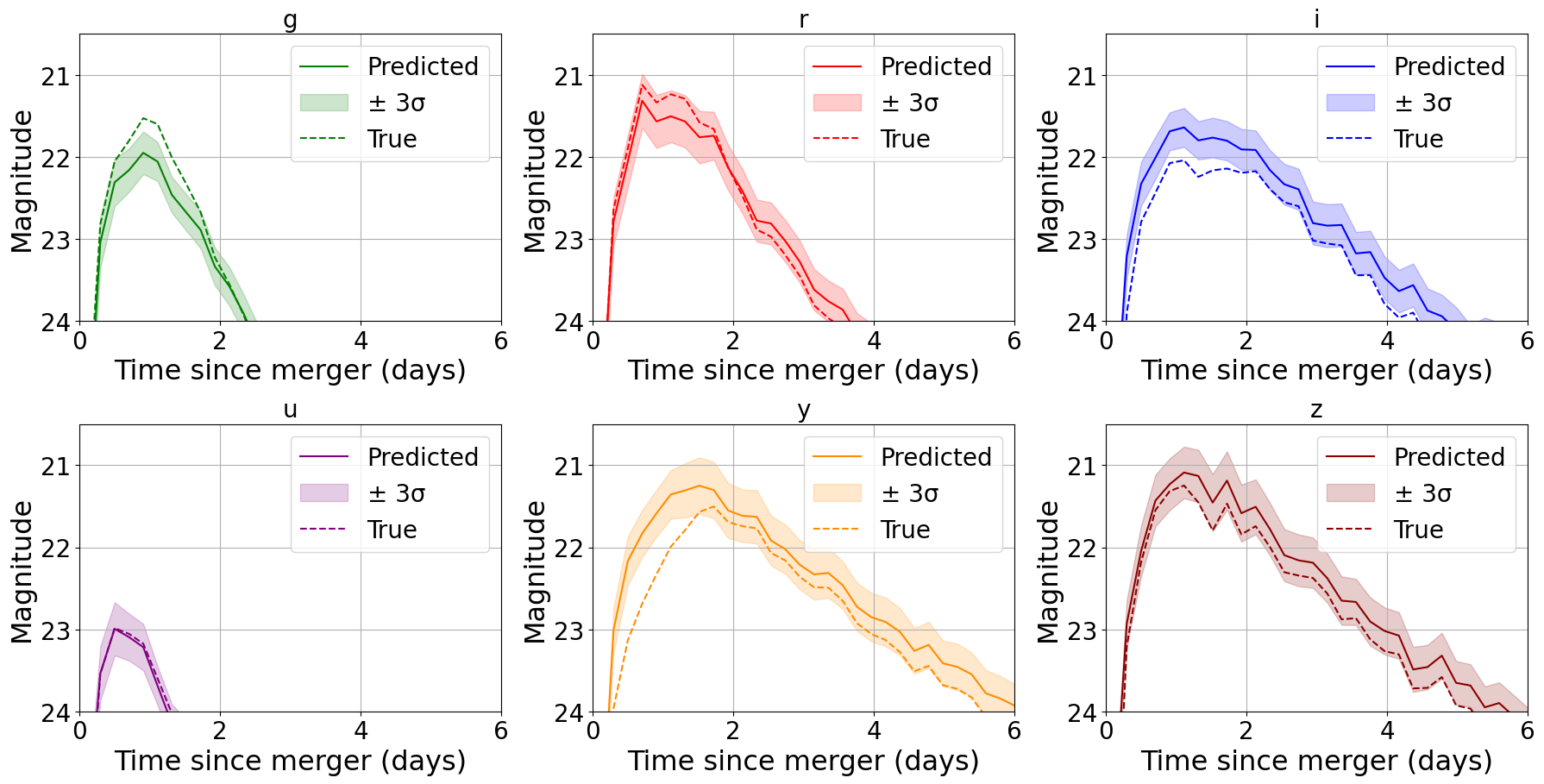}
    \caption{Predicted (solid lines) and ground truth (dashed lines) light curves in Rubin filters. Shaded regions show 3-$\sigma$ uncertainties.}
    \label{fig:corner}
\end{figure*}

\begin{figure*}[t]
    \centering
    \includegraphics[width=0.98\textwidth]{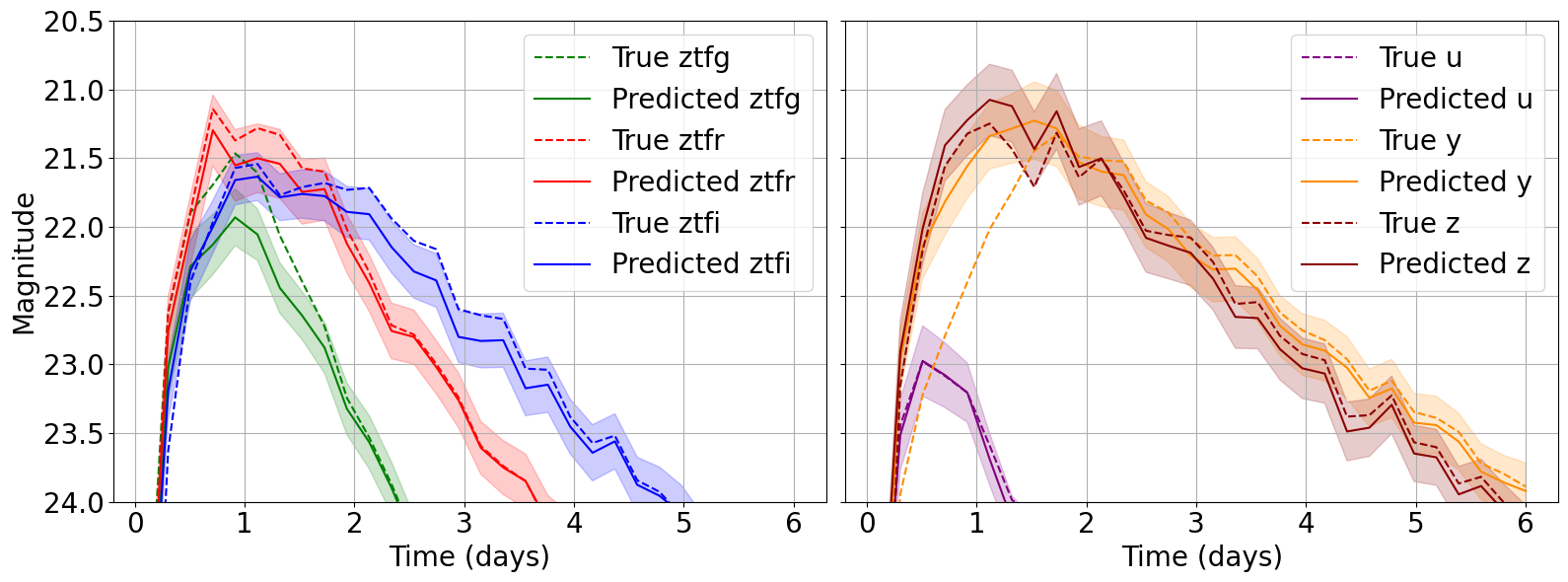}
    \includegraphics[width=0.98\textwidth]{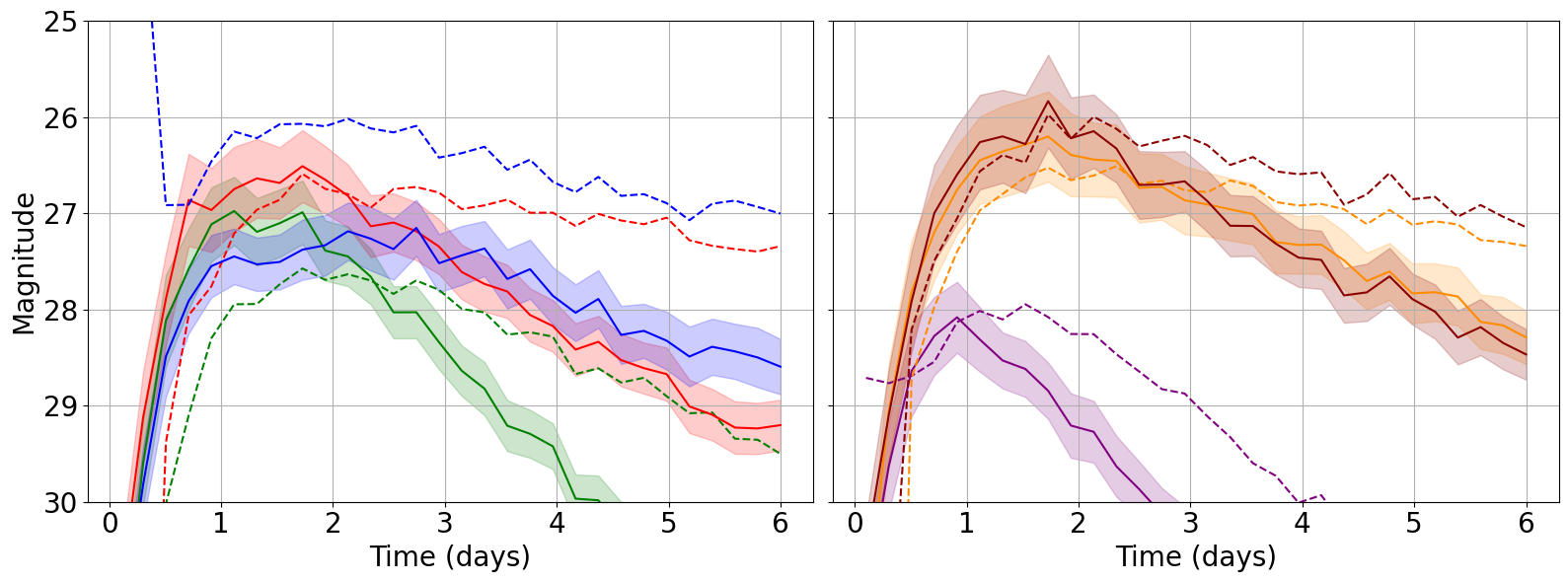}
    \includegraphics[width=0.98\textwidth]{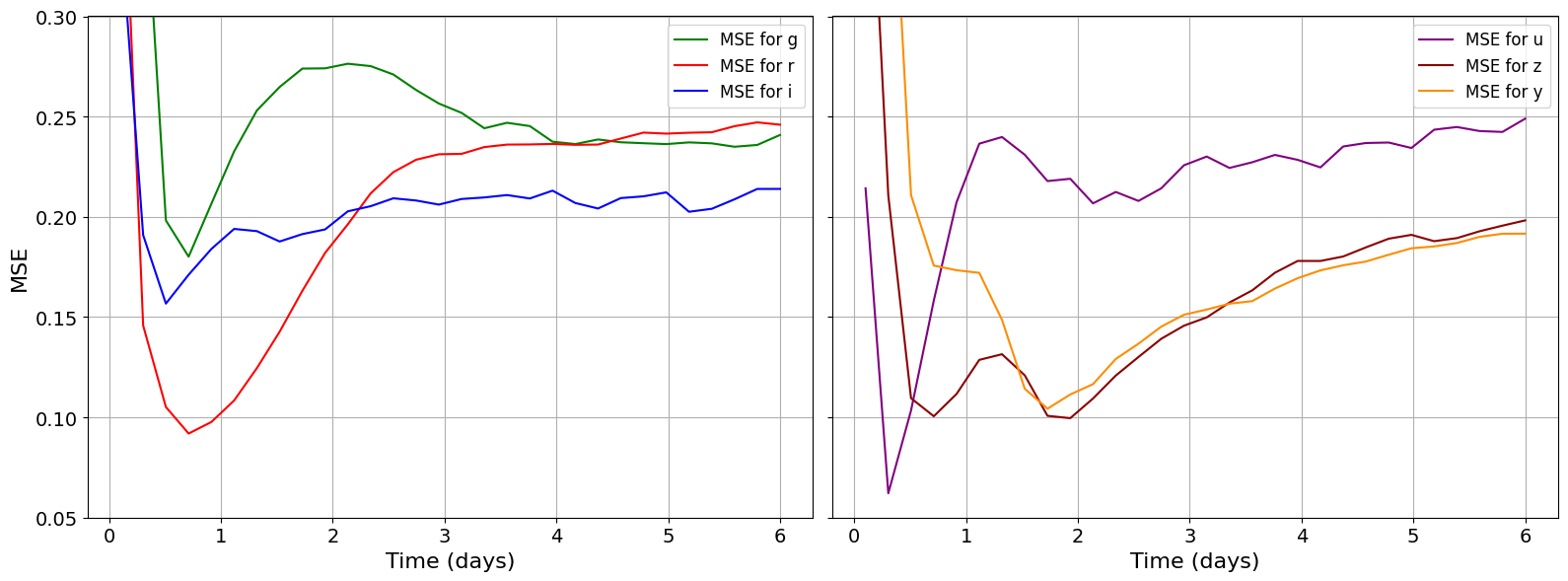}
    \caption{Similar to Figure 3 but showing forecasts in \textit{g}, \textit{r}, \textit{i}, \textit{u}, \textit{y}, and \textit{z} filters, relevant for Rubin. The best-predicted light curve with an MSE of 0.017 is on the top. The worst predicted light curve, with an MSE of 4 in the middle. MSE as a function of KN time is on the bottom.}
    \label{fig:lumfunc}
\end{figure*}

We extended our model to forecast KN light curves across six different filters \textit{u},  \textit{g}, \textit{r}, \textit{i}, \textit{y} and \textit{z} using a more simplified architecture than our fiducial model (see Table 4 in Appendix A) with six input features. This less complex model was adopted because the original configuration struggled to learn effectively and the loss function failed to converge. The multifilter approach aims to enhance the ability of the model to predict light curves across a broader range of wavelengths, including the infrared, in preparation for potential applications with data from the upcoming Rubin Observatory. 

The model achieved an overall mean squared error (MSE) of 0.22 and an $R^2$ score of 0.68 across all six filters on the O5 test data (see Table 3 for more details). Figure 7 shows an example light curve with an MSE of 0.22, where the predicted and actual values match closely. The prediction performance for the optical filters in the O5 run (Rubin filters) is lower than in the O4 run (ZTF filters), primarily due to the smaller dataset size, approximately half as large due to computational limitations. However, the MSE values for the near-infrared filters are lower, indicating the strong predictive capabilities of the model in these bands. This finding is particularly encouraging, as future surveys such as those of the Rubin Observatory will benefit from improved sensitivity in near-infrared bands, which are essential for detecting and characterizing the reddened emission of KNe \citep{Rubin}. The the best-predicted light curves, and the worst-predicted light curves, and the MSE for each Rubin filter as a function of time are shown in Figure 8.

\begin{table}[h]
\centering
\begin{tabular}{|c|c|c|}
\hline
\textbf{Filter} & \textbf{MSE} & \boldmath{$R^2$} \\ \hline
\textit{g}-band & 0.31 & 0.56 \\ \hline
\textit{r}-band & 0.21 & 0.71 \\ \hline
\textit{i}-band & 0.21 & 0.65 \\ \hline
\textit{u}-band & 0.22 & 0.64 \\ \hline
\textit{y}-band & 0.19 & 0.78 \\ \hline
\textit{z}-band & 0.16 & 0.80 \\ \hline
\textbf{Overall} & 0.22 & 0.68 \\ \hline
\end{tabular}
\caption{Model performance per filter on the O5 test data.}
\label{tab:filter_performance}
\end{table}

\subsection{CNN-LSTM Model}

To further improve the performance of our model, we have explored using additional features, such as full skymap information. The proposed model combines CNNs for image processing with bidirectional LSTM networks for sequence prediction. The skymaps are first converted to 3 channel images of size 500x1000. They are normalized using a RobustScaler. This is then processed through a CNN consisting of two convolutional layers with 32 and 64 filters, respectively, each followed by max-pooling layers (see details in Table 5 Appendix B). The resulting features are then flattened and passed through a dense layer. The outputs and the features for our fiducial model are concatenated and passed into a series of bidirectional LSTM layers as previously described.

The CNN-LSTM model's results demonstrate that adding skymap images as extra input features did not improve performance when compared to the LSTM-only model. With a test MSE of 0.24 and an $R^2$ score of 0.80 across ZTF filters, the performance of the model was effectively the same as the LSTM model trained without skymaps. This implies that the skymaps, as used in this approach, did not provide additional information that contributed to better light curve predictions.

Since our fiducial model shows a strong dependence on distance (see Appendix C), the performance of the model can be further improved by exploring different features that could more accurately represent the underlying physical processes of the kilonova event. A promising direction is the addition of features, such as dynamical and wind ejecta, that are directly related to the physics of the event. In the next step, we include the ejecta mass as an additional feature in the model, which significantly improves its performance, achieving an MSE of 0.11 and an $R^2$ score of 0.90 across the ZTF filters, and an MSE of 0.10 and an $R^2$ score of 0.82 across the Rubin filters, using the model described in Appendix A with this additional feature. This shows that the model’s capacity to correctly forecast kilonova light curves may be significantly impacted by using physically relevant features, offering a clear path for future improvements.

\begin{figure*}
    \centering
    \includegraphics[width=0.45\textwidth]{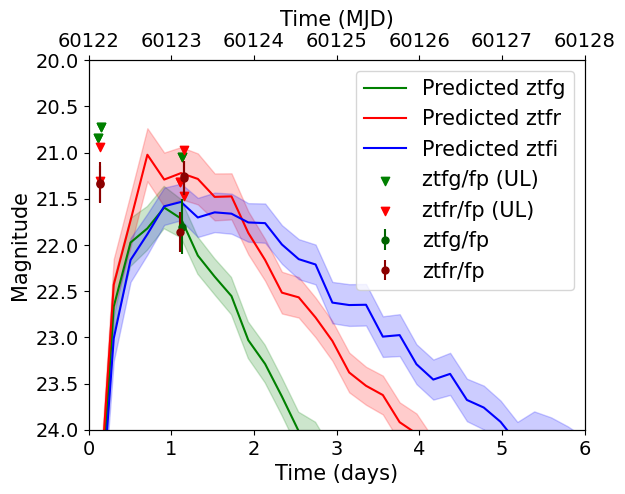}
    \includegraphics[width=0.45\textwidth]{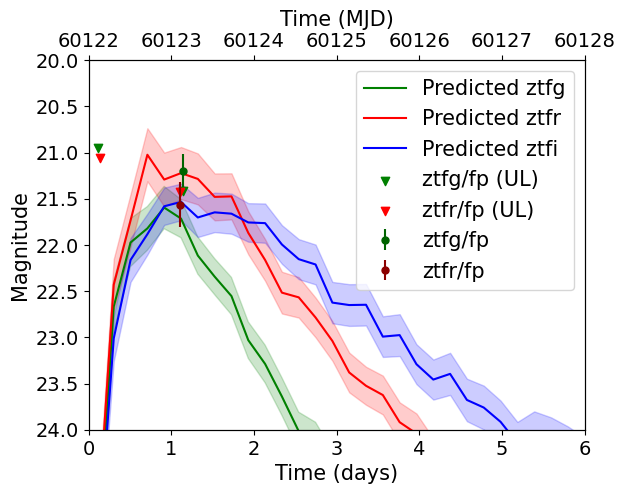}
    \caption{Predicted mean (solid lines) and 3-$\sigma$ uncertainities (shaded regions) of KN light curves associated with S240627c using low latency alert data.}
    \label{fig:lumfunc}
\end{figure*}

\subsection{Comparision to kilonova candidates identified during O4a and b}  

During the IGWN O4a observing run, several GW alerts were followed up by the ZTF collaboration through the Fritz platform \citep{2023Coughlin}. A key candidate was S230627c. We refer to \citet{O4a} for a detailed discussion. We used our production LSTM model (trained with the full dataset: train, validation, and test sets together) to predict the light curve of a potential KN associated with S230627c. Figure 9 compares the forecasts from our model to data from two candidate counterparts. The forecasts generally agree with the detections but disagree during early phases. 

\begin{figure*}
    \centering
    \includegraphics[width=0.45\textwidth]{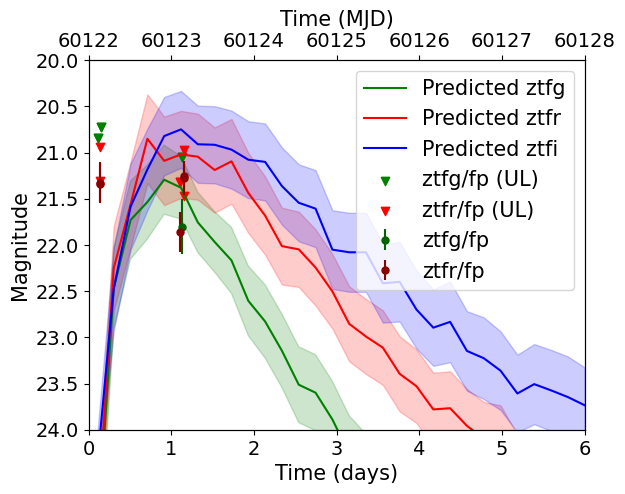}
    \includegraphics[width=0.45\textwidth]{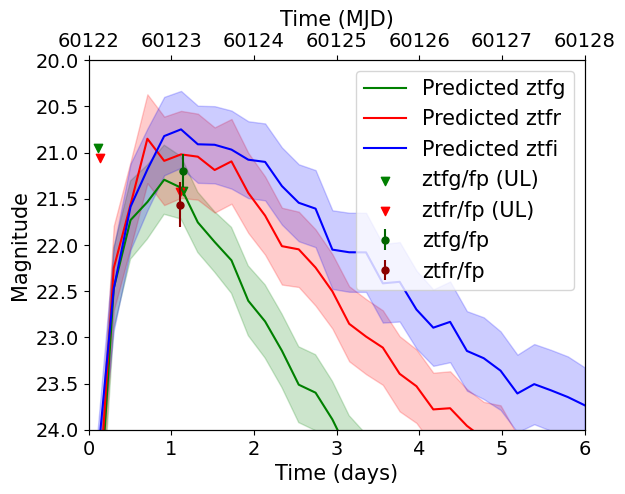}
    \caption{Same as Figure 9 but using a model that uses ejecta mass a feature. While the predictions don't change much in this case, this model overall has a better performance than our fiducial model.}
    \label{fig:lumfunc}
\end{figure*}

Figure 10 shows the predicted light curves from the model that includes ejecta mass as an additional feature. Although the predictions are slightly brighter compared to the model without ejecta information, they remain consistent with observed data points, including during the early phases. The inclusion of dynamical and wind ejecta appears to slightly shift the predicted brightness, but the model still captures the peak and decline phases. This is particularly notable given that the most important feature, ejecta mass, is very small, since this candidate may be a binary black hole (BBH) merger. Although ejecta masses are not available in low-latency, the inclusion of this information can improve the accuracy of light curve forecasts.

None of the candidates is consistent with our forecasts for the 250206dm event \citep{2025Ahumada}. Even after integrating dynamical and wind ejecta as additional features, none of the candidates matched our predictions.

\newpage
\section{Discussion}

In this work, we have created an ML model to predict kilonova light curves using low-latency alert data from the IGWN. This tool can help astronomers quickly estimate whether a kilonova is likely to be visible, how bright the peak will be, and how fast it will fade. It can also help facilitate counterpart searches for low signal-to-noise ratio (high FAR) events, by automatically matching forecasted light curves to transient discoveries. Moreover, once a transient of interest has been identified, it can also help optimize follow-up observations in filter choice and exposure time. This will be especially useful for Rubin discoveries, that will be out of reach for most ground-based resources. To make this tool more useful in real-time, we plan to integrate it into Marchals, e.g. SkyPortal \citep{SkyPorta}, and encourage the LIGO–Virgo–KAGRA (LVK) collaboration to start including these ejecta mass estimates in their public alerts. Coarse-grained mass information, while not yet available in current releases (e.g., \citealt{2024Chaudhary}), is a promising future data product that would further strengthen the connection between GW data and kilonova detectability.

Ideally, a reinforcement learning agent would be used to make follow-up recommendations. An RL agent is necessary because it would take into account the stochastic nature of survey and triggered follow-up data and handle optimization criteria that require the full data to be available for evaluation (e.g. fitting parameters for light curves). Such an agent would maximize the potential of our limited observing resources and is the subject of our future work.

\section*{Acknowledgements}

This research was partly completed on Expanse at the San Diego Supercomputer Center at UC San Diego and Delta at the University of Illinois Urbana-Champaign and its National Center for Supercomputing Applications (Accelerate ACCESS Award\# PHY240341 and \# AST200029). 
N.~Pletskova and N.~Sravan acknowledge support from the National Science Foundation with grant number AST-2307374. N.~Sravan acknowledges support from the Charles E. Kaufman Foundation with a Urania E. Stott grant. M.W.C acknowledges support from the National Science Foundation with grant numbers PHY-2409481, PHY-2308862 and PHY-2117997.

Based on observations obtained with the Samuel Oschin Telescope 48-inch and the 60-inch Telescope at the Palomar Observatory as part of the Zwicky Transient Facility project. ZTF is supported by the National Science Foundation under Grants No. AST-1440341, AST-2034437, and currently Award \#2407588. ZTF receives additional funding from the ZTF partnership. Current members include Caltech, USA; Caltech/IPAC, USA; University of Maryland, USA; University of California, Berkeley, USA; University of Wisconsin at Milwaukee, USA; Cornell University, USA; Drexel University, USA; University of North Carolina at Chapel Hill, USA; Institute of Science and Technology, Austria; National Central University, Taiwan, and OKC, University of Stockholm, Sweden. Operations are conducted by Caltech's Optical Observatory (COO), Caltech/IPAC, and the University of Washington at Seattle, USA.

The ZTF forced-photometry service was funded under the Heising-Simons Foundation grant \#12540303 (PI: Graham).

The Gordon and Betty Moore Foundation, through both the Data-Driven Investigator Program and a dedicated grant, provided critical funding for SkyPortal.

R.W.K. was supported by the French National Research Agency (ANR) under project "Multi-messenger observations of the Transient Sky (MOTS)", no. ANR-22-CE31-0012.

\section*{Software}
All the code used in this study is publicly available on GitHub at \url{https://github.com/np699/Forecast-KN-LC.git}. The repository includes scripts for data preprocessing, model training, light curve forecasting, and uncertainty estimation. It also provides pre-trained models and Jupyter notebooks for reproducing the main results and figures from this paper. The repository contains detailed instructions and code for generating key physical features used in training, including HasNS, HasRemnant, HasMassGap, and $P_{\text{Astro}}$, from simulated or low-latency gravitational wave event parameters. It also includes tools for generating multiband kilonova light curves using the NMMA framework. The training data for the simulated BNS and NSBH merger events is publicly available through Zenodo at \url{https://zenodo.org/records/12696721}. These simulations include low-latency alert parameters.

The model is built in Python using TensorFlow, Keras, NumPy, pandas, scikit-learn, and astropy. All required packages and dependencies are listed in the requirements.txt file in the GitHub repository.

\FloatBarrier
\bibliography{main}
\bibliographystyle{aasjournal}

\appendix

\section{Rubin Architecture}
\label{sec:appendix_model}

\begin{table}[h]
\centering
\renewcommand{\arraystretch}{1.2} 
\begin{tabular}{|c|c|}
\hline
\multicolumn{2}{|c|}{\textbf{Input Features}} \\
\hline
\multicolumn{2}{|c|}{Distance, area(90), HasNS, HasRemnant, HasMassGap, PAstro} \\
\hline

\multicolumn{2}{|c|}{\textbf{Bidirectional LSTM Layers}} \\
\hline
Bidirectional LSTM Layer 1 & 64 units, return\_sequences=True, L1=1e-5, L2=1e-4, max\_norm=3.0 \\
\hline
Dropout & Rate = 0.3 \\
\hline
Bidirectional LSTM Layer 2 & 100 units, ReLU activation, return\_sequences=False, L1=L2=0.001, max\_norm=3.0 \\
\hline
Dropout & Rate = 0.3 \\
\hline

\multicolumn{2}{|c|}{\textbf{Dense Layers}} \\
\hline
Dense Layer 1 & 100 units, ReLU activation \\
\hline
Dropout & Rate = 0.05 \\
\hline
Output Layer & 180 units, Linear activation \\
\hline

\end{tabular}
\caption{Bidirectional LSTM model architecture for forecasting Rubin light curves using 6 input features.}
\label{tab:bilstm_architecture}
\end{table}

\section{CNN-LSTM Model Architecture}
\label{sec:appendix_model}

\begin{table}[h]
\centering
\renewcommand{\arraystretch}{1} 
\begin{tabular}{|c|c|}
\hline
\multicolumn{2}{|c|}{\textbf{Input Features}} \\ 
\hline
\multicolumn{2}{|c|}{Skymap images, Distance, area(90), HasNS, HasRemnant, HasMassGap, PAstro} \\ 
\hline
\multicolumn{2}{|c|}{\textbf{CNN for Skymap Images}} \\ 
\hline
Conv2D Layer 1 & 32 filters, (3x3), ReLU, padding='same' \\ 
\hline
MaxPooling2D & Pool size (2x2) \\ 
\hline
Conv2D Layer 2 & 64 filters, (3x3), ReLU, padding='same' \\ 
\hline
MaxPooling2D & Pool size (2x2) \\ 
\hline
Dense Layer & 128 units, ReLU activation \\ 
\hline

\multicolumn{2}{|c|}{\textbf{Feature Concatenation}} \\ 
\hline
Concatenation & CNN output + Feature vector \\ 
\hline

\multicolumn{2}{|c|}{\textbf{Bidirectional LSTM Layers}} \\ 
\hline
Details & Same as in Table~\ref{tab:hyperparameters} \\ 
\hline

\multicolumn{2}{|c|}{\textbf{Output Layer}} \\ 
\hline
Units & 90 \\ 
\hline
Activation Function & Linear \\ 
\hline

\end{tabular}
\caption{CNN-LSTM model architecture with multimodal inputs (skymap images and features). LSTM details are the same as in Table~\ref{tab:hyperparameters}.}
\label{tab:cnn_lstm_architecture}
\end{table}

\newpage
\section{Feature Importance}
\label{sec:appendix_model}

\begin{figure*}[h]
    \centering
    \includegraphics[width=0.48\textwidth]{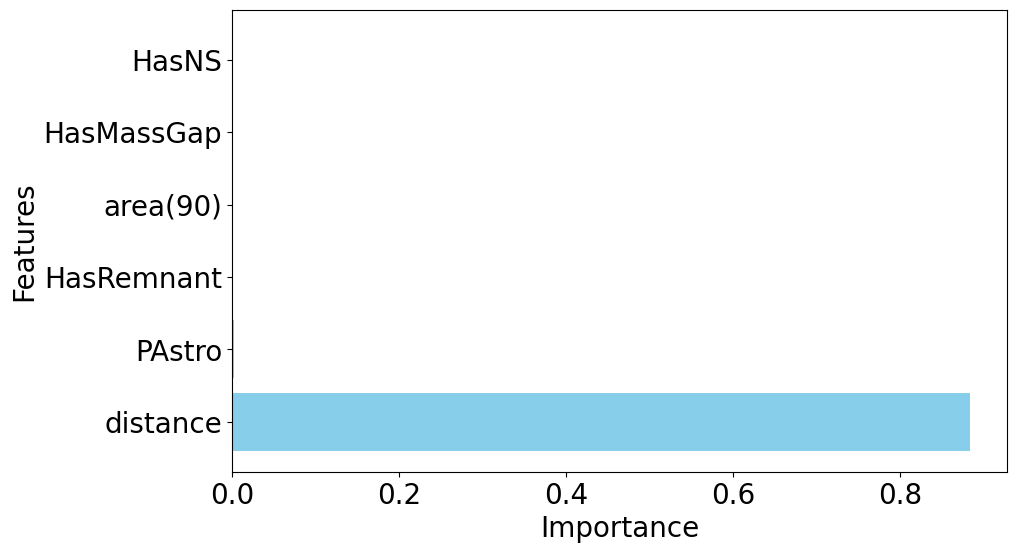}
    \includegraphics[width=0.48\textwidth]{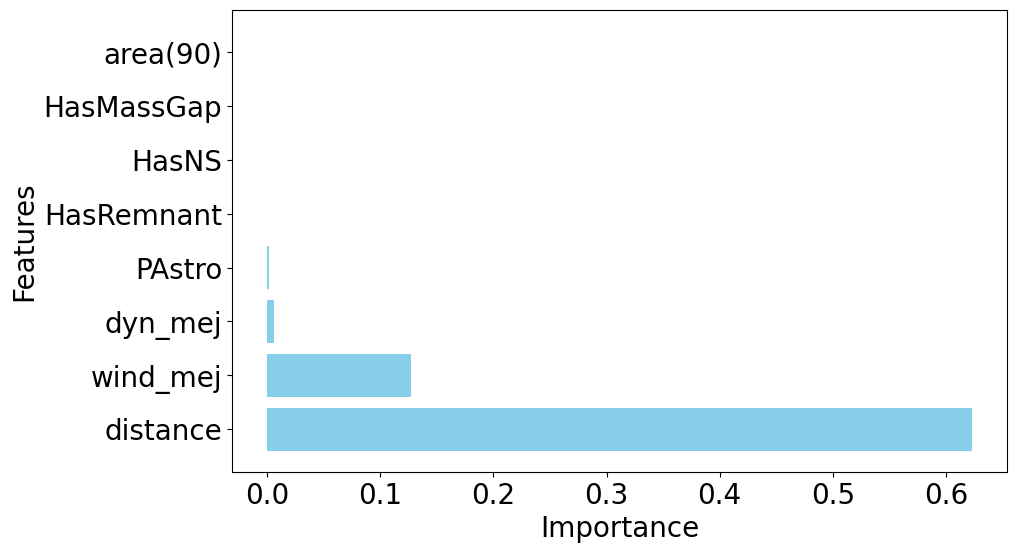}
    \caption{Feature importance of the fiducial model without ejecta mass (left) and with ejecta mass included (right)}
    \label{fig:lumfunc}
\end{figure*}

\section{Peak magnitude vs FAR}
\label{sec:appendix_model}
\begin{figure*}[h]
    \centering
    \includegraphics[width=0.98\textwidth]{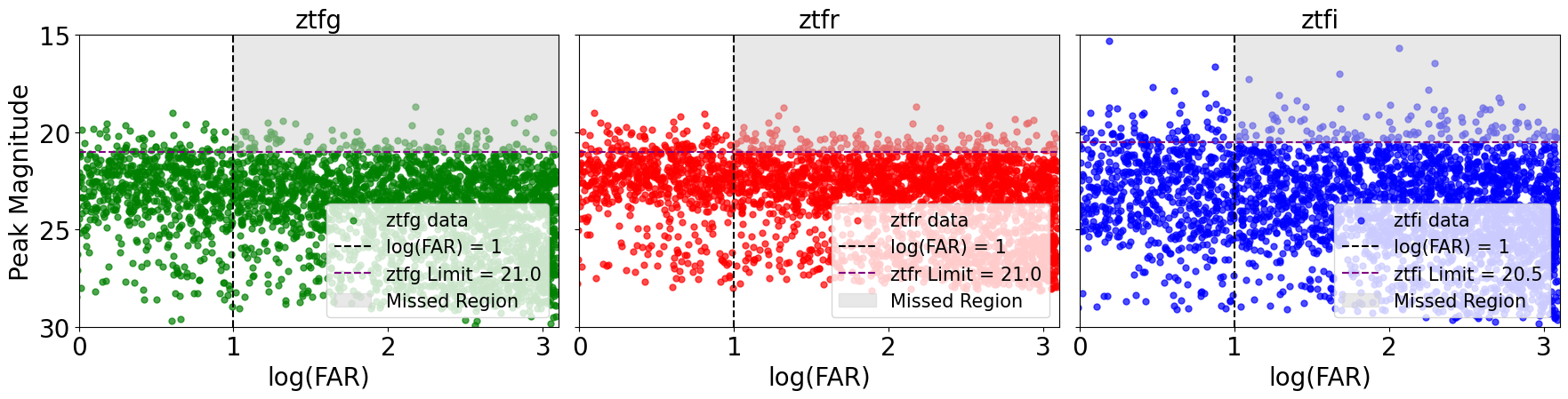}
    \caption{Peak magnitude vs FAR for O4 CBC detections. The purple line indicates the ZTF magnitude limit and the black line delineates FAR = 22. KNe associated with CBCs in the gray region are currently missed by follow-up programs using this FAR cut.}
    \label{fig:lumfunc}
\end{figure*}

\end{document}